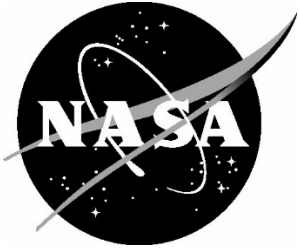

# Application of Metric-Based Mesh Adaptation to Hypersonic Aerothermal Simulations Using US3D

*Dirk Ekelschot*
*Analytical Mechanics Associates (AMA) at NASA Ames Research Center, Moffett Field, CA*

# NASA STI Program Report Series

Since its founding, NASA has been dedicated to the advancement of aeronautics and space science. The NASA scientific and technical information (STI) program plays a key part in helping NASA maintain this important role.

The NASA STI program operates under the auspices of the Agency Chief Information Officer. It collects, organizes, provides for archiving, and disseminates NASA's STI. The NASA STI program provides access to the NTRS Registered and its public interface, the NASA Technical Reports Server, thus providing one of the largest collections of aeronautical and space science STI in the world. Results are published in both non-NASA channels and by NASA in the NASA STI Report Series, which includes the following report types:

- TECHNICAL PUBLICATION. Reports of completed research or a major significant phase of research that present the results of NASA Programs and include extensive data or theoretical analysis. Includes compilations of significant scientific and technical data and information deemed to be of continuing reference value. NASA counterpart of peer-reviewed formal professional papers but has less stringent limitations on manuscript length and extent of graphic presentations.

- TECHNICAL MEMORANDUM.
  Scientific and technical findings that are preliminary or of specialized interest,
  e.g., quick release reports, working papers, and bibliographies that contain minimal annotation. Does not contain extensive analysis.

- CONTRACTOR REPORT. Scientific and technical findings by NASA-sponsored contractors and grantees.

- CONFERENCE PUBLICATION.
  Collected papers from scientific and technical conferences, symposia, seminars, or other meetings sponsored or co-sponsored by NASA.

- SPECIAL PUBLICATION. Scientific, technical, or historical information from NASA programs, projects, and missions, often concerned with subjects having substantial public interest.

- TECHNICAL TRANSLATION.
  English-language translations of foreign scientific and technical material pertinent to NASA's mission.

Specialized services also include organizing and publishing research results, distributing specialized research announcements and feeds, providing information desk and personal search support, and enabling data exchange services.

For more information about the NASA STI program, see the following:

- Access the NASA STI program home page at http://www.sti.nasa.gov

.

NASA/TM–20250011734

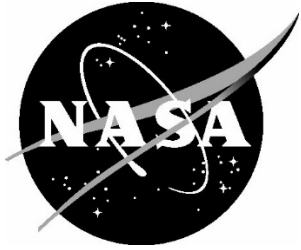

# Application of Metric-Based Mesh Adaptation to Hypersonic Aerothermal Simulations Using US3D

*Dirk Ekelschot*
*Analytical Mechanics Associates (AMA) at NASA Ames Research Center, Moffett Field, CA*

# Application of Metric-Based Mesh Adaptation to Hypersonic Aerothermal Simulations Using US3D

D. Ekelschot*
*Analytical Mechanics Associates, Moffett Field, NASA Ames Research Center*

The main goal of this paper is to demonstrate the application of metric-based mesh adaptation to real gas problems and highlight the benefits particularly when complex geometries are considered. We use the Hessian of the temperature solution as an indicator to dictate where the mesh needs refinement or coarsening. In the context of hypersonic flow simulations, these methods are not widely adopted since unstructured meshes often result in poor surface heating predictions. The present work aims to demonstrate the great flexibility metric-based mesh adaptation provides when it comes to predicting complex flow features while still maintaining comparable surface heating predictions. We consider two test cases: (a) a supersonic flow over a hemisphere and show that comparable surface heating is obtained by applying mesh adaptation and by employing hexahedra instead of prisms in the boundary layer mesh; (b) we consider a more realistic test case of a hypersonic flow of a $CO_2$-$N_2$ mixture past a 70° sphere cone atmospheric entry capsule. For the second test case, similar surface heating predictions are obtained compared to more conventional block structured DPLR simulations. Furthermore, for the adapted unstructured simulations, the geometries of the eight Reaction Control System (RCS) jet on the backshell were taken into account. This highlights the ability of these methods to deal with complex geometries that are typically out of reach for block structured approaches.

## I. Introduction

The current state-of-the-art Computational Fluid Dynamics (CFD) codes that are used to simulate hypersonic flows, like DPLR [1] and US3D [2], typically employ a second-order finite volume discretization to predict the surface heating on the forebody as long as the mesh consists of hexahedral elements that are properly aligned with the bow shock in front of the entry vehicle. For simple geometries, the generation of a shock-aligned meshes is relatively straightforward. However, the process becomes more cumbersome once the geometry becomes more complex and/or the flow exhibits complex flow features.

Recently, interest increased in capturing the wake dynamics and studying the influence of the wake on the back shell heating for atmospheric entry vehicles. Typically, the surface heating that occurs at the forebody is significantly higher than on the aft body. However, the uncertainties in heat transfer predictions on the backshell remain significantly higher on the backshell compared to the forebody due to the chaotic dynamics of the wake flow. The flow is unsteady and typically exhibits flow features like shock wave-boundary layer interactions, shear layers and shock-shock interaction to name a few. Feature based mesh adaptation would be desirable in this case but block structured hexahedral meshes are typically not flexible enough when it comes to solution driven mesh refinement.

Alternatively, unstructured anisotropic mesh adaptation can be used to improve the simulation resolution in the vicinity of these complex flow phenomena. The research presented in the present work focuses on anisotropic mesh adaptation for mixed unstructured meshes that consist of a fixed boundary layer mesh near the wall that can consists of prisms, hexahedra and pyramids and unstructured anisotropic tetrahedra in the remainder of the volume. Employing solution-based mesh adaptation strategies makes it possible to automate and simplify the time-consuming mesh generation step. However, tetrahedral meshes are known to provide inaccurate heating predictions in the context of hypersonic flow simulations particularly when they are not aligned with the bow shock. Metric-based mesh adaptation can potentially solve this problem.

The goal here is to investigate the feasibility of metric-based mesh adaptation in the context of hypersonic flow simulations using the recently developed Metric-Informed Mesh Improvement Capability (MIMIC) [3, 4] in combination with the unstructured flow solver US3D. MIMIC enables metric-based mesh adaptation for mixed element meshes in the context of compressible flow simulations using the compressible flow solver US3D [2]. This work is a follow-on to [3, 4] where MIMIC was applied to supersonic and transonic atmospheric entry problems. In this work, the application of MIMIC is extended to hypersonic flow problems where real gas effects are considered. In previous work, the surface mesh was tessellated using unstructured triangles, which were extruded into prismatic layers to form the boundary layer mesh. The current work extends this capability to support quadrilateral surface elements that can be extruded into hexahedral layers. This paper shows examples of this new utility and compares results surface heat transfer results between using a prismatic or a hexahedral boundary layer mesh. This enhancement enables engineers to combine unstructured and locally block-structured meshing strategies on boundaries, as demonstrated in the final example of this paper. The remainder of the volume is then tessellated using unstructured tetrahedra which are ultimately adapted to improve the CFD resolution. We illustrate the flexibility of the mesh adaptation strategy for problems with multiple complex physical phenomena like shock waves and wake effects for example.

---

*Aerothermodynamicst, Analytical Mechanics Associates at NASA Ames Research Center

The outline of this paper is as follows: First, a brief overview of the relevant literature is given in section II. Second, an overview of the numerical approach and the employed anisotropic mesh adaptation strategy is given in section III. This is followed by two examples in section IV. The first example demonstates the application of a prismatic versus a hexahedral boundary layer mesh in the context of anisotropic mesh adaptation for the canonical case of supersonic flow past a hemisphere. In the second example, we apply MIMIC to study hypersonic flow past a 70° sphere cone atmospheric entry capsule at peak heating conditions. Finally, a brief discussion of the conclusions is provided in section V.

## II. Literature review

Previous studies have shown that improper alignment of the mesh with the strong bow shock in front of the fore body results in poor quality of surface heat transfer predictions [5, 6]. Consequently, the current state-of-the-art hypersonic flow simulations rely on block structured meshes that consist of hexahedra shaped elements which have proven to provide accurate surface heating predictions. Hypersonic simulation softwares like DPLR [1] and US3D [2] tailor the outer boundary of the computational domain to lie as closely as possible to the bow shock in order to reduce the number of elements in the domain and lower the computational cost. Meshes that contain tetrahedra in the volume are known to perform poorly when it comes to simulating hypersonic flows. The bow shock introduces significant amounts of numerical entropy that convects downstream to the surface of the geometry and ultimately pollutes the surface heat flux predictions at the forebody [5]. Recently, more work has been done on improving numerical schemes in order to improve surface heat flux predictions with unstructured meshes [7, 8]. However, mesh allignment with shocks is still required to improve surface heat flux predictions. Examples of applying best practices from structured mesh generation to unstructured meshes are given by McCloud [9] and Tang [10]. Recently, Nastac et al[11, 12] have employed metric-based mesh adaptation with a fixed prismatic boundary layer mesh similarly to what is being shown here in this paper. They have used this capability to predict surface heat transfer for realistic three-dimensional atmospheric entry geometries [11, 12]. They show that similar levels of accuracy can be achieved with shock adapted tetrahedra compared to manually crafted hexahedral meshes. The advantage here being the option to automate the mesh generation/adaptation process and improve the resolution in regions where other interesting flow phenomena happen in the wake of the geometry for example. The work by Morgado et al.[13] is another example where anisotropic mesh adaptation was applied in the context of surface heat flux predictions and a similar level of accuracy was achieved compared to approach that rely on structured meshes.

## III. Computational Framework

The aim here is to use metric-based mesh adaptation in the context of hypersonic flow simulations. In this section, we describe both the advanced compressible flow solver US3D and the Metric-Informed Mesh Improvement Capability (MIMIC) which handles the anisotropic mesh adaptation step based on a provided US3D mesh and a compatible solution field.

### A. US3D flow solver

In the present work, we use the advanced compressible flow solver US3D [2] to simulate flow that consists of a mixture of multiple species. US3D is a cell-centered finite volume code that is able to simulate flow on unstructured meshes and is essentially the successor of the NASA DPLR CFD code [1]. US3D uses a Modified Steger-Warming flux vector splitting approach [14]. The Kinetic Energy Consistent (KEC) finite volume scheme, proposed by Subbareddy et al. [15], is available in US3D. However, the modified Steger-Warming flux vector splitting approach is employed for all the simulations shown in the paper. US3D is able to simulate compressible flow that is considered to be a perfect gas flow but also a multi-species mixture of reacting gases. In this work, for the simulations that consider a reacting mixture of gasses, the vibrational, electronic and free electron energy modes are grouped together and the thermodynamic properties are computed using the NASA Lewis database. Furthermore, US3D is able to model turbulence using either the Reynolds Averaged Navier-Stokes (RANS) equations or using an Improved Delayed Detached-Eddy (IDDES) approach [16]. However, in this work we do not employ a turbulence model.

**B. Metric-Informed Mesh Improvement Capability (MIMIC)**

The Metric-Informed Mesh Improvement Capability (MIMIC) has been recently developed in order to enable metric-based mesh adaptation for advanced compressible flows [3, 4]. MIMIC has been initially applied to simulate transonic and supersonic flow where a perfect gas was considered. In this work, we consider a multi-species mixture of reacting flow and aim to establish a work flow that can be used to obtain accurate heat flux predictions using unstructured meshes. An additional contribution of this work is to demonstrate that MIMIC is able to handle hexahedra next to prisms in the boundary layer mesh. Before, the user was able to only use triangles to tesselate the surface mesh and extrude those triangles into layers of prisms in order to generate a structured boundary layer mesh. This feature has now been extended such that the user can generate a quadrilateral surface mesh and extrude those quadrilateral elements into layers of hexahedra near the geometry.

*1. Computing the metric tensor field*

MIMIC consists of three main components. First, it computes the metric tensor field based on a provided mesh and initial solution field. Anisotropic mesh adaptation aims to align the mesh with a metric tensor field that prescribes the size and directionality of the local interpolation error [3, 17–26]. This metric tensor field is essentially a local error indicator that depends on the second-order gradients (Hessian) of the provided solution field. MIMIC uses the weighted least-squares gradient reconstruction method to compute the Hessian. By default, MIMIC reconstructs the first-order gradients first and then iterates again to compute the second-order gradients so that the Hessian is computed. Alternatively, MIMIC also has the option to extend the scheme and reconstruct the second-order gradients directly as shown in [4]. The tetrahedra are separated from the other element types such as prisms, pyramids, hexahedra. Both the volume that consists of tetrahedra and the boundary layer mesh are treated as separated parts of the domain and partitioned individually in order to load balance and make sure that all the data is equaly distributed in memory. The tetrahedra are adapted based on the computed metric tensor field while the other element types typically define a fixed boundary layer mesh near the geometry. Consequentely, MIMIC solely computes the metric tensor field for the tetrahedra to limit computational cost.

*2. Employ ParMMG to adapt the mesh*

The second component involves passing the metric tensor field to the open-source anisotropic mesh adaptation Application Programming Interface (API) called ParMMG [27, 28]. ParMMG computes a new mesh that is aligned with the underlying metric. ParMMG is the parallel version of the open-source mesh adaptation library, MMG3D, and it performs anisotropic Delaunay mesh adaptation in parallel based on the provided metric tensor field [29]. The adaptation is performed solely on tetrahedral shaped elements. At the start of the adaptation procedure, the mesh is partitioned and each partition carries out a local adaptation keeping the shared faces between partitions fixed. ParMMG makes sure that the vertices on those shared interfaces are accounted for as well by iteratively re-partitioning and re-adapting the mesh. Hence, after each partition is done adapting, a re-partitioning of the global mesh is carried out such that the vertices and faces that were shared are now internal and therefore taken into account during the next adaptation phase. Typically four to six re-partitioning iterations are used for each adaptation iteration in order to ensure that all vertices in the mesh are accounted for.

*3. Input/Output*

The final and third component involves writing the distributed adapted mesh into a single mesh file (HDF5) that is compatible with the US3D flow solver. In the current version, the MIMIC code is able to output the adapted mesh into CFD General Notation System (CGNS) format. This allows the user to load the adapted mesh output by MIMIC into a commercial mesh generation software like Pointwise [30]. The adapted mesh can then be output into a format that is compatible with a different flow solver that employs different discretization schemes.

## C. MIMIC workflow

MIMIC relies on a first flow solution that is computed on a coarse hybrid mesh. This initial coarse hybrid mesh is referred to as the base mesh and is typically genereated using an external mesh software tool like Pointwise [30]. A surface mesh is generated and this surface mesh is extruded into a structured boundary layer mesh. Farfield domains are typically specified by a box and the volume between the outer layer of the boundary layer mesh and the farfield domain is tesselated using ustructured tetrahedra that will ultimately be adapted. The current implementation allows the user to define a hybrid boundary layer mesh that consists of prisms, hexahedra and pyramids, and that remains fixed during the adaptation procedure [3]. A first error estimate is derived based on this coarse representation of the solution and the mesh. The user defines the number of adaptation iterations, $N$, and that mesh is adapted $N$ times based on the temperature solution field. In the context of the simulations presented in this paper, a first-order flow solution is obtained by solving the RANS equations. This provides us with a first estimate of where the shock waves and shear layers are located. After that, the user can decide to either add additional adaptation iterations based on various simulation modelling choices available in US3D. In the case of unsteady flows, statistical data needs to be collected when running the US3D flow simulation so that the mesh can be adapted based on the mean temperature field. Furthermore, MIMIC allows the user to refine the mesh isotropically in the wake based on the Turbulent Kinetic Energy (TKE) [3].

# IV. Results

In this paper, we will consider both a perfect gas flow and a multi-species mixture of reacting gases. One of the goals of this paper is to establish a comparison between accuracy in heat flux predictions obtained using anisotropic mesh adaptation versus heat flux predictions that are obtained using a conventional block structured hexahedral mesh.

## A. Supersonic flow past a hemisphere

As a first example, we consider supersonic flow ($U = 1000\ m/s$, $\rho_\infty = 0.01\ kg/m^3$, $T_\infty = 200\ K$) past a hemisphere with an isothermal wall at $T_w = 500\ K$. We consider air that behaves as a perfect gas that has a specific heat ratio of $\gamma = 1.4$ and a Prandtl number of $Pr = 0.72$. This example aims to determine the sensitivity of surface heat transfer results on the local mesh configuration (prisms versus hexahedra near the surface of the geometry).

Next to that, the thickness of the boundary layer mesh is varied in order to see how this affects the surface heat transfer predictions as well. As a result, we consider two different mesh configurations. The first mesh has an unstructured triangular tessellation at the surface of the hemisphere and the second mesh configuration has an unstructured quadrilateral surface tesselation. Meshes that have a triangular surface tesselation are denoted by $\mathcal{T}_p$ and meshes that have a quadrilateral surface tesselation are referred to as $\mathcal{T}_h$. The surface mesh is extruded for both $\mathcal{T}_p$ and $\mathcal{T}_h$ into either 80 or 110 prismatic layers or hexahedral layers respectively. An initial wall normal spacing of $1.0 \times 10^{-5}$ $m$ is used and a growth rate of $1.05$ for this extrusion operation is considered. In the discussion of the results, the number of prismatic or hexahedral layers is added as a subscript. Furthermore, multiple adaptation iterations are run for each of these meshes and the iteration number is indicated by a superscript. As an example, we refer to the fifth adaptation iteration for a mesh that has a triangular surface tesselation and 80 prismatic layers as $\mathcal{T}_{p,80}^5$ while a base mesh that has a quadrilateral surface tesselation with 110 layers of hexahedra as $\mathcal{T}_{h,110}^0$. For both $\mathcal{T}_{p,N}^0$ and $\mathcal{T}_{h,N}^0$, the remainder of the domain is tessellated into isotropic tetrahedra.

First, we consider a mesh topology that has a triangular unstructured surface mesh. An impression of the initial base mesh ($\mathcal{T}_{p,80}^0$) is shown in Figure 1a. This mesh contains 80 prismatic layers near the wall and the temperature and surface heat transfer contours are shown in Figure 1b and Figure 1c respectively. Figure 1c shows a highly distorted surface heat flux distribution which is typically obtained when using a coarse mesh that consists of unstructured tetrahedra. Figure 1b illustrates the presence of a strong bow shock in the temperature contours. The initial coarse solution shown in Figure 1b, is used to adapt the mesh as described in section III.B. This is done iteratively, and for all cases shown in this example, five adaptation iterations are performed based on the Hessian of the temperature solution. Figure 1d shows the mesh that is obtained after five adaptation iterations and a clear clustering of highly stretched elements are shown in the vicinity of the shock. Figure 1e also shows a significant improvement in shock definition. Figure 1f depicts the surface heat flux distribution that is calculated using $\mathcal{T}_{p,80}^5$. Comparing Figure 1f with Figure 1c shows a clear improvement and the peak heating appears to be more centered towards the stagnation point but the contours do still show a relatively noisy surface heat flux distribution.

In addition to these results, the same surface mesh is considered but now the surface mesh is extruded into 110 prismatic layers. The same initial wall normal spacing of $1.0 \times 10^{-5}$ $m$ and growth rate of $1.05$ is used. An example of

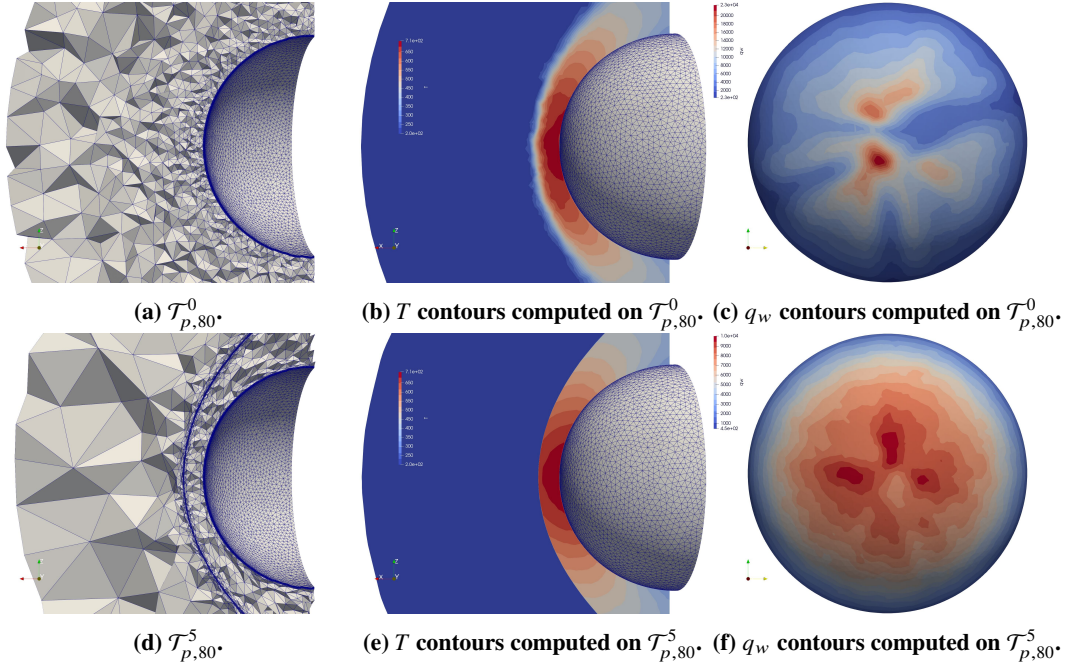

**(a)** $\mathcal{T}_{p,80}^0$.    **(b)** $T$ contours computed on $\mathcal{T}_{p,80}^0$.    **(c)** $q_w$ contours computed on $\mathcal{T}_{p,80}^0$.

**(d)** $\mathcal{T}_{p,80}^5$.    **(e)** $T$ contours computed on $\mathcal{T}_{p,80}^5$.    **(f)** $q_w$ contours computed on $\mathcal{T}_{p,80}^5$.

**Fig. 1** Comparing temperature and surface heat transfer contours between $\mathcal{T}_{p,80}^0$ and $\mathcal{T}_{p,80}^5$.

the initial base mesh ($\mathcal{T}_{p,110}^0$) is shown in Figure 2a. The corresponding temperature and surface heat flux contours are shown in Figure 2b and Figure 2c. Once again, the temperature contours illustrate a diffusive solution due to the coarseness of the mesh and the heat flux contours show a similar distorted heat flux distribution as shown in Figure 1c. The temperature and surface heat transfer solution is significantly improved after applying five adaptation iterations as shown in Figures 2d-2f. The surface heat flux distribution shown in Figure 2f appears to be more smooth compared to the surface heat flux shown in Figure 1f. Even though same levels of adaptation are applied for both $\mathcal{T}_{p,80}$ and $\mathcal{T}_{p,110}$, the surface heat flux solution appears to be sensitive to the thickness of the structured boundary layer mesh near the surface of the geometry which indicates that the boundary layer is not properly resolved in for the $\mathcal{T}_{p,80}$ meshes.

In the remainder of this section, we consider a quadrilateral surface mesh that is extruded into either 80 or 110 layers of hexahedra. Similarly to the prismatic cases that were discussed before, the same initial wall normal spacing of $1.0 \times 10^{-5}$ is used and the hexahedra are extruded with a growth rate of 1.05. First a boundary layer mesh that consists of 80 layers of hexahedra near the wall is considered. The surface heat flux plotted in Figure 3a shows streaks particularly near the stagnation point. Similarly to the previous cases for the prismatic meshes, we adapt the mesh five times. The improved surface heat flux distribution is shown in Figure 3d. The surface heat flux distribution is significantly improved but similarly to Figure 1d oscillations near the the stagnation point are observed. Finally, the quadrilateral surface mesh is extruded into 110 layers of hexahedra and the corresponding initial mesh, temperature contours and heat flux contours are shown in Figure 4a- 4c while the final mesh, temperature contours and heat flux contours are shown in Figure 4d- 4f. Figure 4f shows a significant improvement in surface heat flux distribution compared to Figure 4c. As expected, the heat flux distribution shown in Figure 4f appears to be more symmetric than the heat flux distribution shown in Figure 3f. In general, it is concluded that there is a strong sensitivity observed of the calculated surface heat flux with respect to the height of the structured boundary layer mesh. Next to that the hexahedral boundary layer mesh appears to result in smoother symmetric surface heat flux distribution compared to the prismatic boundary layer mesh as shown in Figure 5.

Figure 5 compares the pressure and heat flux profiles for both the two different prismatic and hexahedral boundary layer mesh topologies and highlights the effect of mesh adaptation. The pressure errorbars that are shown in Figure 5a-5c are currently chosen somewhat arbitrarily to be ±2% of the maximum pressure. Similarly, the heat flux errorbars, shown in Figure 5d-5f, are chosen to be ±7.5% of the maximum heat flux. These errorbars are included here to essentially quantify what acceptable heat flux profile are that are computed on unstructured adapted meshes. Figures 5a and 5b illustrate that the pressure profile is accurately recovered once mesh adaptation is applied for all boundary layer mesh topologies. Figure 5c shows a comparison of the pressure profile after five adaptation iterations for both $\mathcal{T}_{p,110}^5$ and

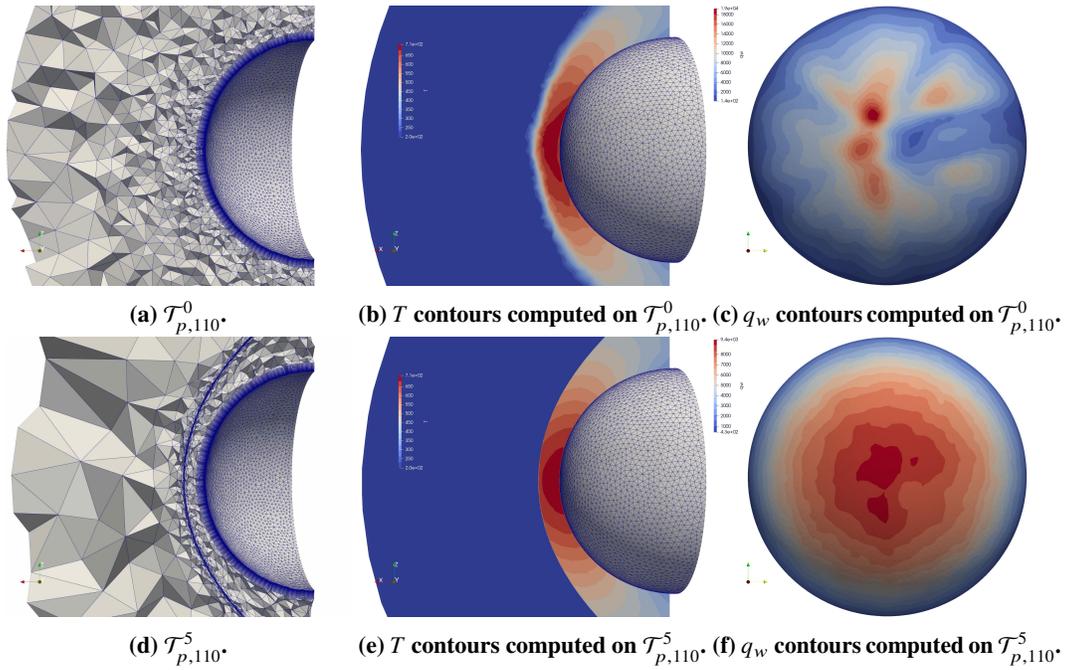

**(a)** $\mathcal{T}_{p,110}^{0}$.  **(b)** $T$ contours computed on $\mathcal{T}_{p,110}^{0}$.  **(c)** $q_w$ contours computed on $\mathcal{T}_{p,110}^{0}$.

**(d)** $\mathcal{T}_{p,110}^{5}$.  **(e)** $T$ contours computed on $\mathcal{T}_{p,110}^{5}$.  **(f)** $q_w$ contours computed on $\mathcal{T}_{p,110}^{5}$.

**Fig. 2** Comparing temperature and surface heat transfer contours between $M_{p,110}^{i}$ and $\mathcal{T}_{p,110}^{5}$.

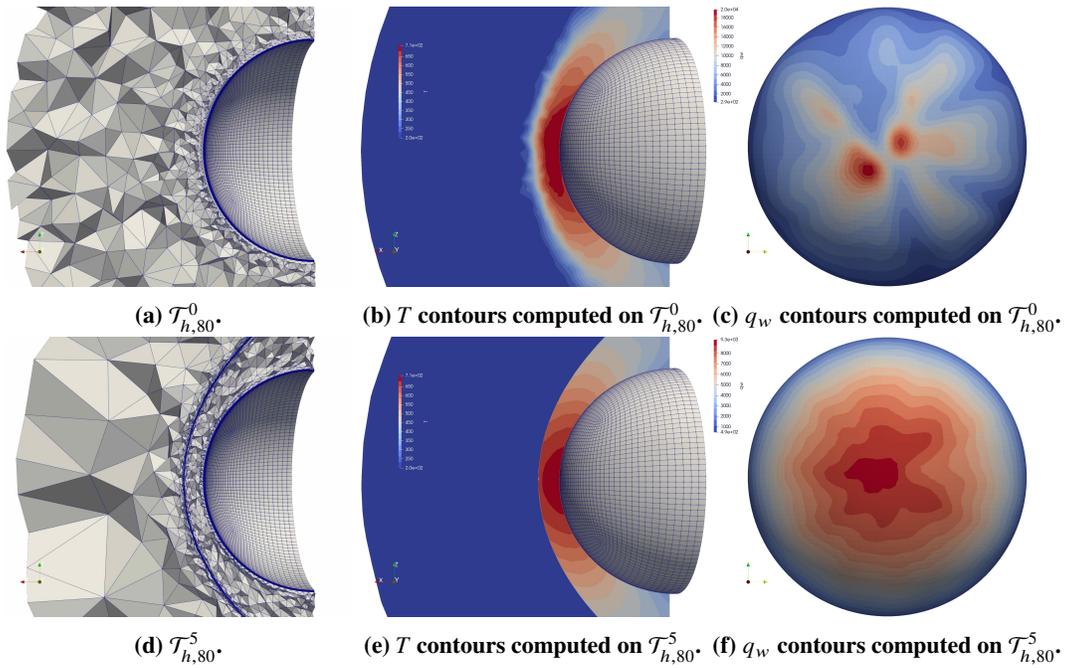

**(a)** $\mathcal{T}_{h,80}^{0}$.  **(b)** $T$ contours computed on $\mathcal{T}_{h,80}^{0}$.  **(c)** $q_w$ contours computed on $\mathcal{T}_{h,80}^{0}$.

**(d)** $\mathcal{T}_{h,80}^{5}$.  **(e)** $T$ contours computed on $\mathcal{T}_{h,80}^{5}$.  **(f)** $q_w$ contours computed on $\mathcal{T}_{h,80}^{5}$.

**Fig. 3** Comparing temperature and surface heat transfer contours between $\mathcal{T}_{h,80}^{0}$ and $\mathcal{T}_{h,80}^{5}$.

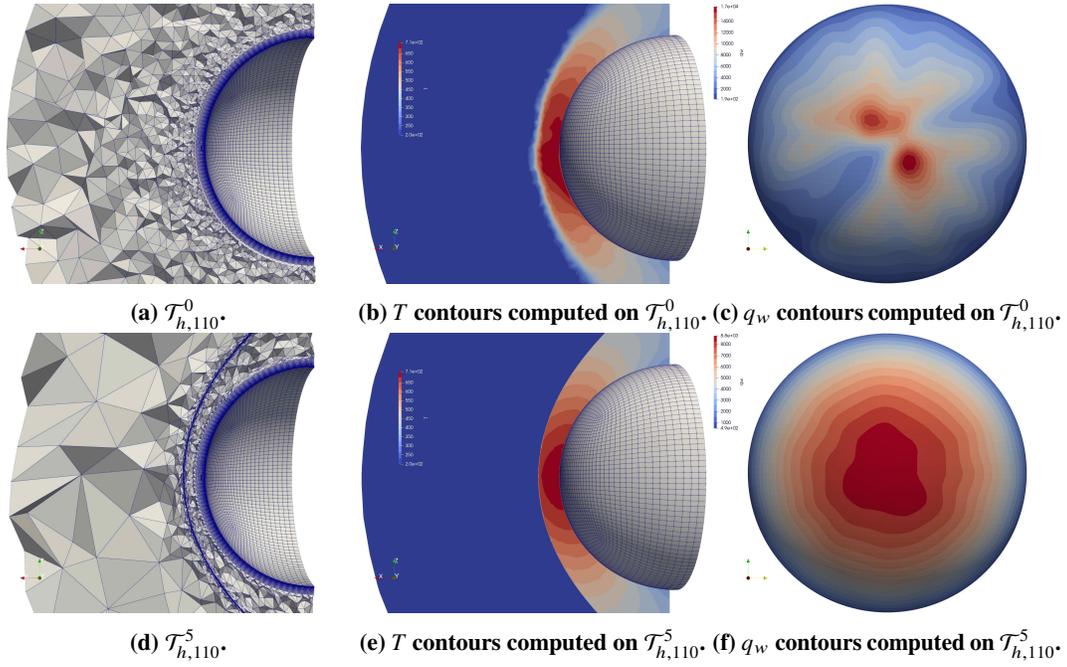

(a) $\mathcal{T}_{h,110}^0$.

(b) $T$ contours computed on $\mathcal{T}_{h,110}^0$.

(c) $q_w$ contours computed on $\mathcal{T}_{h,110}^0$.

(d) $\mathcal{T}_{h,110}^5$.

(e) $T$ contours computed on $\mathcal{T}_{h,110}^5$.

(f) $q_w$ contours computed on $\mathcal{T}_{h,110}^5$.

**Fig. 4** Comparing temperature and surface heat transfer contours between $\mathcal{T}_{h,110}^0$ and $\mathcal{T}_{h,110}^5$.

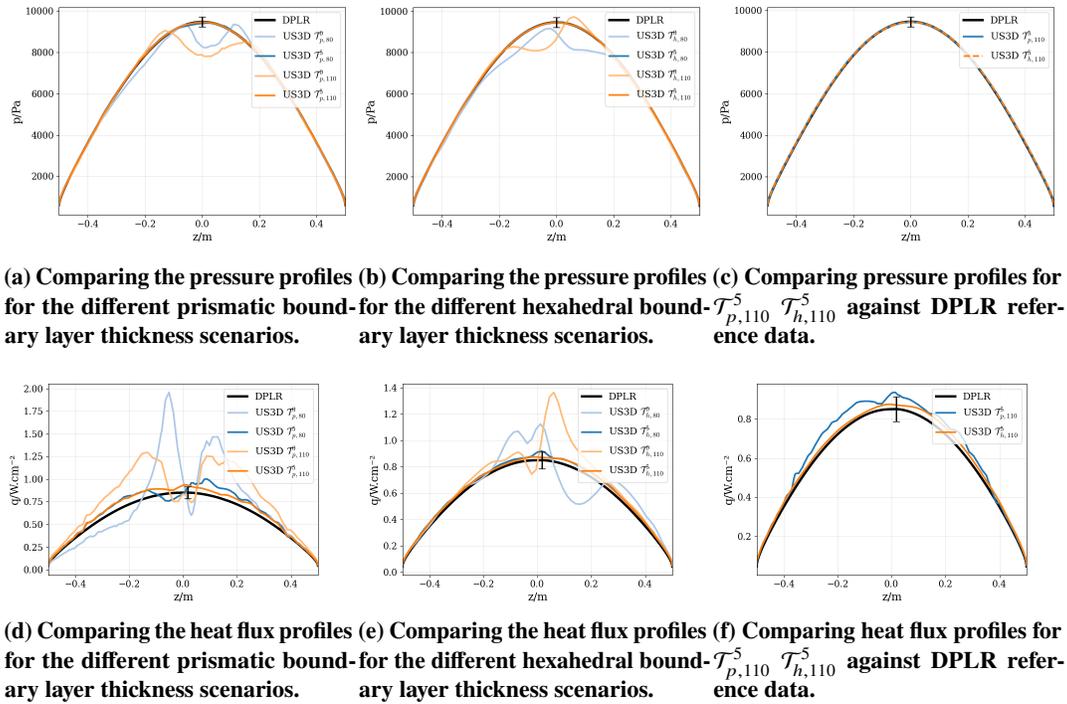

(a) Comparing the pressure profiles for the different prismatic boundary layer thickness scenarios.

(b) Comparing the pressure profiles for the different hexahedral boundary layer thickness scenarios.

(c) Comparing pressure profiles for $\mathcal{T}_{p,110}^5$ $\mathcal{T}_{h,110}^5$ against DPLR reference data.

(d) Comparing the heat flux profiles for the different prismatic boundary layer thickness scenarios.

(e) Comparing the heat flux profiles for the different hexahedral boundary layer thickness scenarios.

(f) Comparing heat flux profiles for $\mathcal{T}_{p,110}^5$ $\mathcal{T}_{h,110}^5$ against DPLR reference data.

**Fig. 5** Comparing pressure and surface heat transfer profiles for the different boundary mesh topologies and mesh adaptation iterations against DPLR reference data.

$\mathcal{T}^5_{h,110}$ against DPLR reference data and the lines fall on top of each other. Figures 5d and 5e show the heat flux profiles for the prismatic and hexahedral boundary layer mesh topology respectively. After five adaptation iterations, the heat flux predictions improve particularly for $\mathcal{T}^5_{h,110}$. Figure 5f compares the heat flux profiles computed on $\mathcal{T}^5_{p,110}$ and $\mathcal{T}^5_{h,110}$ with the DPLR reference data and shows that using a hexahedral boundary layer mesh in combination with sufficient mesh adaptation results in a heat flux profile prediction that lies within the previously stated error bounds.

## B. Hypersonic flow past a 70° sphere cone entry capsule with Reaction Control System (RCS) jets.

As a second example, we consider a hypersonic flow ($U_\infty = 4522\ m/s$, $\rho_\infty = 1.71E - 03\ kg/m^3$, $T_\infty = 167.4\ K$, $M_\infty = 21.3$) of reacting gasses past a 70° sphere cone entry capsule. These are free-stream conditions at which the Sample Return Lander (SRL) capsule expects to experience peak heating during its Martian atmospheric entry phase. For this example, the eight Reaction Control System (RCS) jets on the capsule's backshell are taken into account as well. For the purpose of demonstration, we consider just flow ingestion into the idle nozzles. We do not consider any additional inflow from any of the RCS jets so no plume wake interaction is simulated at this time. The main goal here is to demonstrate and highlight the necessity of employing adaptive unstructured meshing strategies when complex geometries are considered. We consider a gas mixture that replicate a Martian atmosphere and that has the following free-stream mass fractions:

$$\begin{aligned}\mathbf{y}_\infty &= [y_{CO_2}, y_{CO}, y_{C_2}, y_{N_2}, y_{O_2}, y_{NO}, y_{CN}, y_C, y_N, y_O] \\ &= [0.9709, 0.0, 0.0, 0.0291, 0.0, 0.0, 0.0, 0.0, 0.0, 0.0]\end{aligned}$$

A two temperature model is used with vibration-electronic energy relaxation enabled using the NASA Lewis database. The flow enters the computational domain under an angle of attack of $\alpha = 20°$. The 70° sphere cone entry vehicle has a diameter of approximately $D = 4.6m$ and an overview of the geometry is shown in Figure 6.

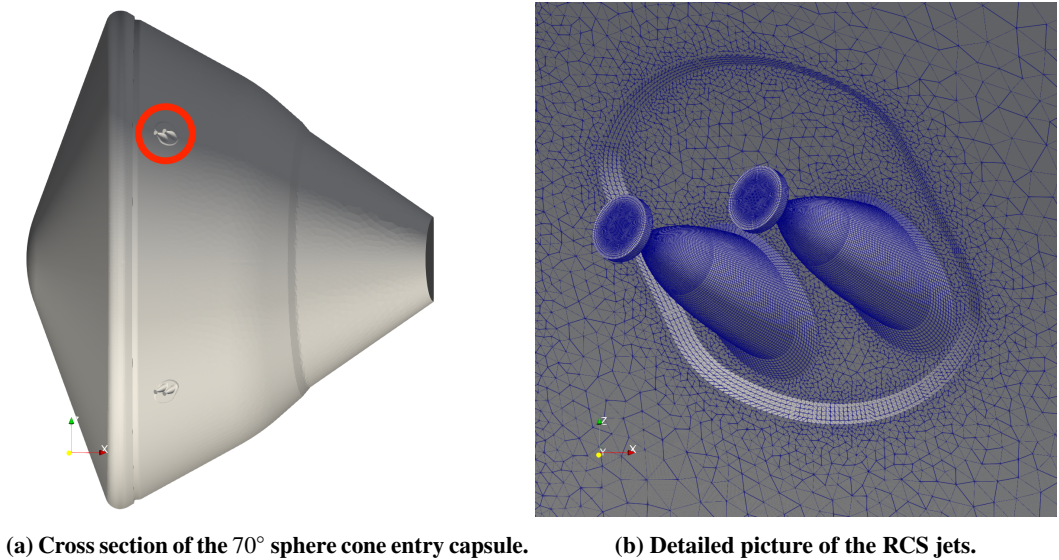

(a) Cross section of the 70° sphere cone entry capsule.  (b) Detailed picture of the RCS jets.

**Fig. 6  Overview of the 70° entry capsule with RCS jets on the backshell.**

Figure 6a shows a cross section of the entry capsule, taken at the symmetry plane of the vehicle, which illustrates the location of the RCS jets. In total, eight RCS jets are present on the backshell and a detailed picture of two of the RCS jets is shown in Figure 6b.

For the initial unstructured base mesh, a mixture of quadrilateral and triangular surface elements are considered. Each RCS nozzle has approximately $2.00 \times 10^5$ quadrilateral surface elements. The forebody consists of approximately $3.00 \times 10^5$ quadrilateral elements and the aftbody has approximately $1.30 \times 10^5$ triangular and $8.50 \times 10^4$ quadrilateral elements near the back facing step (see Figure 11a). These surface elements are extruded in normal direction into 63 layers of hexahedra and prisms so that a structured boundary layer mesh is generated near the surface. The wall normal spacing is $5.5 \times 10^{-5}\ m$ and the growth rate of the hexahedra/prisms in normal direction is 1.085. The remainder of the domain is tessellated using isotropic tetrahedra as shown in Figure 7a. As a result, most of the mesh budget goes to the boundary layer mesh, since the surface triangles and quadrilateral elements are grown out into 63 layers of surface elements. This process results in approximately $8.20 \times 10^6$ prisms, $1.84 \times 10^7$ hexahedra and $2.90 \times 10^5$ pyramids. The pyramids here are required in order to transition from the hexahedra elements to tetrahedra elements. Consequently, the boundary layer mesh contains approximately $2.69 \times 10^7$ elements which is 75% of the total mesh budget of the initial base mesh.

Once the base mesh is set up, a statistically converged solution is computed that is used to start the iterative mesh adaptation process. In this example, the temperature field is used to drive the mesh adaptation cycles. Five adaptation iterations were performed and a summary of the mesh statistics are shown in the table below:

Table 1   Mesh statistics for each adaptation iteration

|  | $\mathcal{T}_0$ (base) | $\mathcal{T}_1$ | $\mathcal{T}_2$ | $\mathcal{T}_3$ | $\mathcal{T}_4$ | $\mathcal{T}_5$ |
| --- | --- | --- | --- | --- | --- | --- |
| $N_{hex} + N_{pri} + N_{pyr}$ | $2.69 \times 10^7$ | $2.69 \times 10^7$ | $2.69 \times 10^7$ | $2.69 \times 10^7$ | $2.69 \times 10^7$ | $2.69 \times 10^7$ |
| $N_{tet}$ | $8.00 \times 10^6$ | $1.22 \times 10^7$ | $1.17 \times 10^7$ | $1.64 \times 10^7$ | $2.86 \times 10^7$ | $3.85 \times 10^7$ |
| $N_{tot}$ | $3.49 \times 10^7$ | $3.91 \times 10^7$ | $3.86 \times 10^7$ | $4.33 \times 10^7$ | $5.55 \times 10^7$ | $6.54 \times 10^7$ |
| $N_{vrt}$ | $2.41 \times 10^7$ | $2.49 \times 10^7$ | $2.48 \times 10^7$ | $2.56 \times 10^7$ | $2.76 \times 10^7$ | $2.93 \times 10^7$ |

*1. Comparing results for the various adaptation iterations*

An initial RANS simulation is performed using the Menter's Shear Stress Transport (SST) turbulence model. Once the steady state solution is obtained, the simulation is restarted without an active turbulence model and a new statistically converged solution is obtained. The initial flow solution together with the initial base mesh are shown in Figure 7.

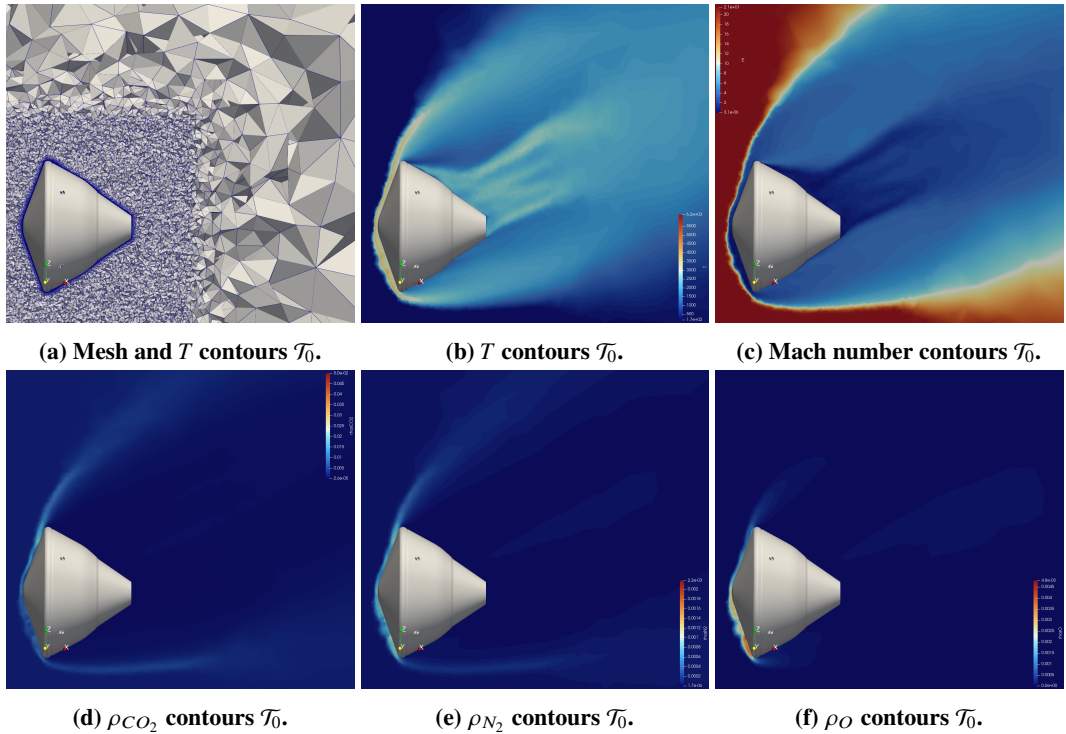

(a) Mesh and $T$ contours $\mathcal{T}_0$.  (b) $T$ contours $\mathcal{T}_0$.  (c) Mach number contours $\mathcal{T}_0$.

(d) $\rho_{CO_2}$ contours $\mathcal{T}_0$.  (e) $\rho_{N_2}$ contours $\mathcal{T}_0$.  (f) $\rho_O$ contours $\mathcal{T}_0$.

Fig. 7   Initial solution for hypersonic flow past a $70°$ cone entry capsule on $\mathcal{T}_0$.

Figure 7a shows that the mesh is locally finer near the body but becomes very coarse further away from the geometry. Figures 7b-7f illustrate very diffusive solution contours but there is a clear bow shock present upstream of the vehicle and some initial wake features can be observed based on the temperature and Mach contours. Figures 7d-7f show that the approximations of the local mass fractions particularly in the vicinity of the bow shock are very noisy due to the coarseness and missalignment of the local mesh. At first, the solution shown in Figure 7b was used to adapt the mesh. During the initial runs it was noticed that utilizing the second-order inviscid flux discretization directly results in strong carbuncle effects which would interfere with the adaptation process. An example of this can be seen in Figure 7c and Figure 7f where the shock near the stagnation line demonstrates signs of weak carbuncle. As a results, a more diffusive initial solution was obtained by switching to the first-order inviscid flux implementation available in US3D. Consequently, a more damped initial solution is obtained which is more suitable to drive the initial mesh adaptation steps.

The solution calculated on $\mathcal{T}_0$ is interpolated onto the new mesh, $\mathcal{T}_1$, and the simulation is restarted until a new statistically converged solution is obtained. This is repeated five times in total. For the first two adaptation iterations, the first-order inviscid flux implementation is employed to drive the mesh adaptation. The solution contours that were obtained after the second adaptation iteration are shown in Figure 8. Figure 8a shows clear allignment of the mesh with

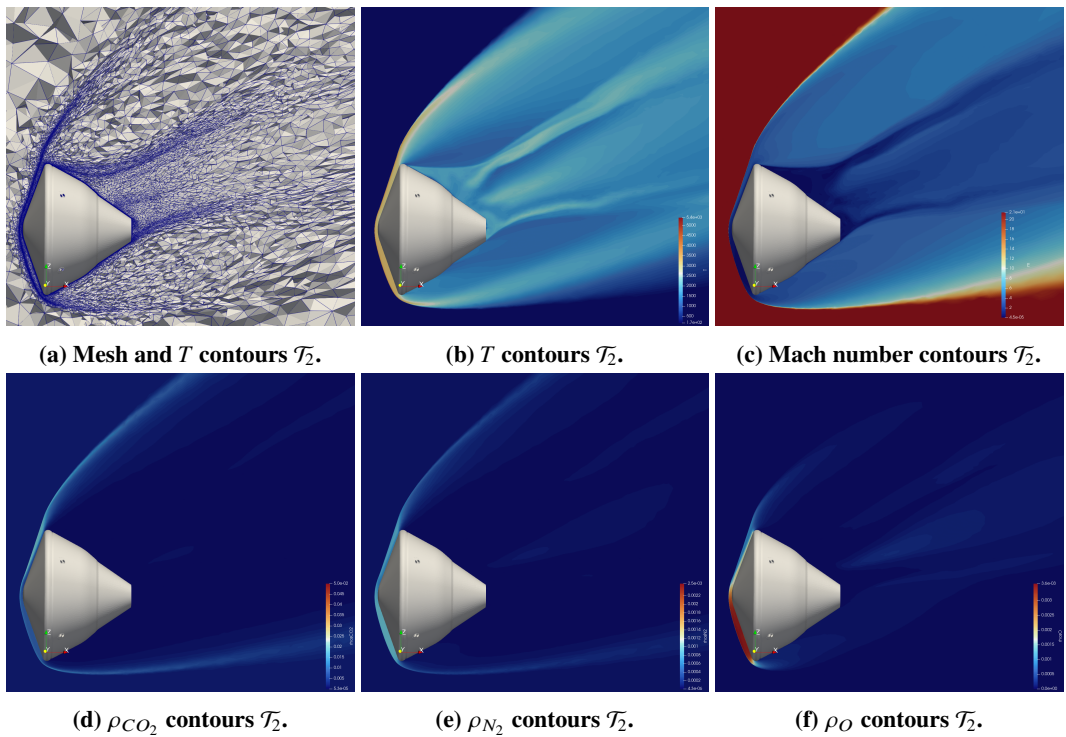

(a) Mesh and $T$ contours $\mathcal{T}_2$.  (b) $T$ contours $\mathcal{T}_2$.  (c) Mach number contours $\mathcal{T}_2$.

(d) $\rho_{CO_2}$ contours $\mathcal{T}_2$.  (e) $\rho_{N_2}$ contours $\mathcal{T}_2$.  (f) $\rho_O$ contours $\mathcal{T}_2$.

**Fig. 8   Solution for hypersonic flow past a $70°$ cone entry capsule on $\mathcal{T}_2$.**

the bow shock and shear layers that eminate from the shoulder of the vehicle. The temperature and Mach contours shown in Figure 8b and 8c illustrate a more defined representation of the bow shock and local wake features. Furthermore, Figures 8d-8f show that smooth mass fraction contours are obtained and the perturbutations shown in Figures7d-7f have been drastically reduced. This can be ascribed to the local allignment and refinement of the mesh.

The solution contours that correspond to the final fifth adaptation iteration are shown in Figure 9. Figure 9a illustrates further refinement near the bow shock and in the wake. Notably, the mesh seems to refine in the vicinity of the shear layers and as a result, interesting unsteady interactions between the recirculation bubble and the shear layer are observed. This becomes clearly visible when looking at the temperarture contours shown in Figure 9b where the upper shear layer appears to break up into three seperate shear layers. The flow solution is now highly unsteady in the wake. The local mesh refinement captures smaller spatial scales that would be impractical to capture with a single-block structured meshes required for DPLR simulations. There has been a drastic improvement in resolution in the vicinity of the bow shock and the wake region when we compare Figure 9 with Figure 7.

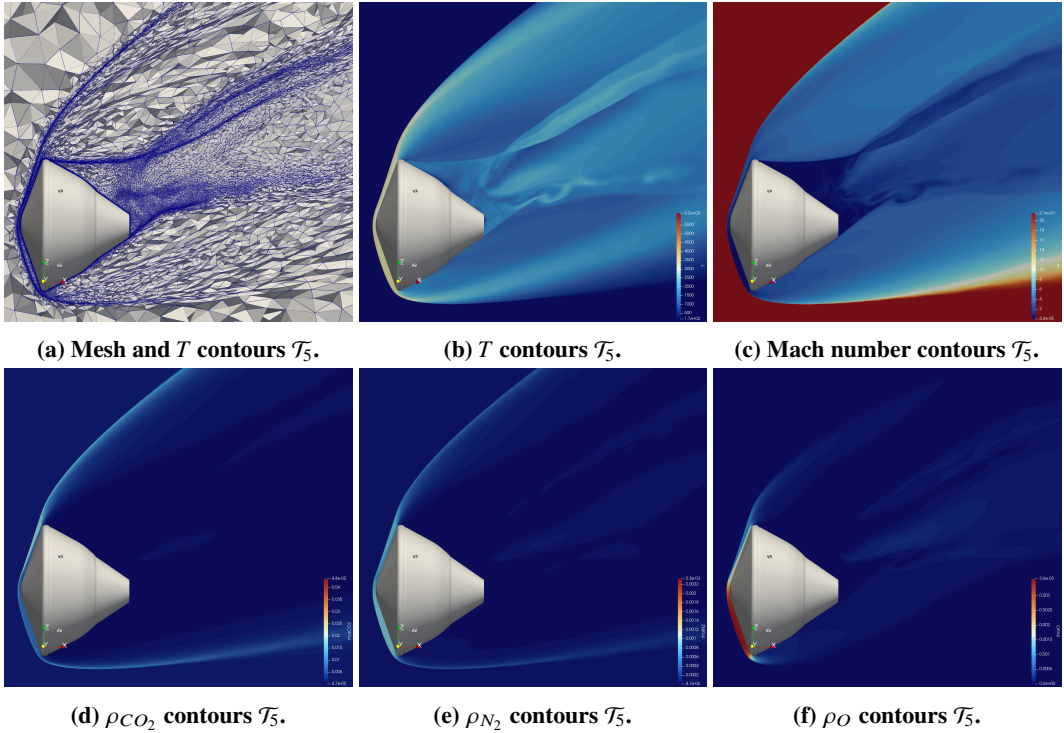

Fig. 9 Solution for hypersonic flow past a $70°$ cone entry capsule on $\mathcal{T}_5$.

*2. The effect of mesh adaptation on surface heat flux predictions on the forebody*

Ultimately, one of the main quantities of interest is the predicted convective heat flux at the surface. Figure 10 shows the surface heat flux on the forebody for the initial base simulation (Figure 10a) and the five adaptation iterations (Figure 10b- 10f). Figure 10a shows large streaks in the surface heat flux distribution on the forebody. These streaks are ascribed to numerical errors that are introduced particularly in the vicinity of the bow shock. These numerical errors arise from the coarse, unstructured mesh topology, which fails to capture the strong gradients in flow quantities. As a result, the errors are convected downstream onto the forebody. Streaks are still clearly visible after the first adaptation iteration but the distribution became less noisy. Furthermore, the heat flux near the lower shoulder seems to reduce which corresponds to areas where the shock standoff distance is larger. There is a noticeable improvement in surface heat flux contours after the second adaptation iteration (see Figure 10c) where the mesh becomes more aligned with the strong bow shock. Note that only the tetrahedra in the volume are adapted and the local boundary layer mesh near the body remains fixed during the adaptation process. Figure 10c and Figure 10d demonstrate further improvement but there are still irregularities present in the surface heat flux contours particularly near the stagnation region. The second-order inviscid flux discretization was employed after the third adaptation iteration and this results in further improvement in surface heat flux when we compare Figure 10d with Figure 10c. The strong persistent streak that was present in Figure 10a- 10c has now almost vanished. The surface heat flux contours that correspond to the fourth and fifth adaptation iteration are shown in Figure 10e and Figure 10f. After the fifth adaptation iteration, the surface heat flux is nearly symmetic and most of the mesh induced streaks have vanished.

*3. Qualitative comparisons of adaptive US3D results with a DPLR reference solution*

First, a detail of the surface mesh for the US3D simulations is compared with a representative surface mesh that is used for a DPLR simulation in Figure 11a and Figure 11b respectively. DPLR requires the mesh to consist solely of structured hexahedral elements. Figure 11b shows that geometric feature refinement on the surfaces results in localized mesh clustering or nested refinement due to local block adjacent mesh sizing requirements. Note, that the mesh shown in Figure 11b does not take into account the geometry of the RCS jets that is shown in Figure 6b and considers the geometry to be smooth. In contrast, the unstructured approach allows the user to fully incorporate the RCS thruster geometry in the simulation as shown in Figure 6b.

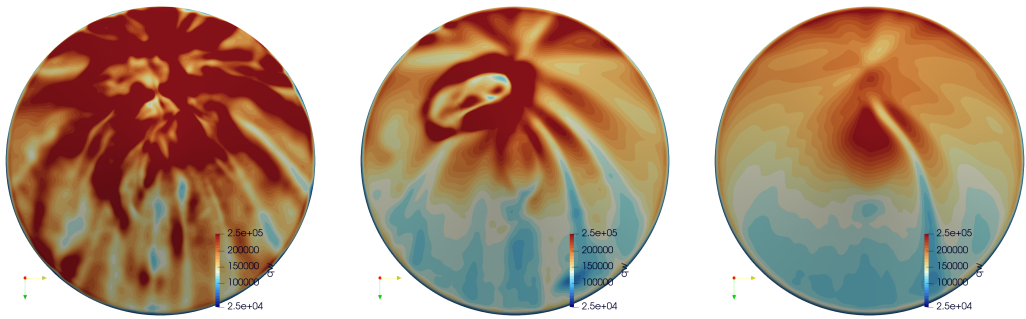
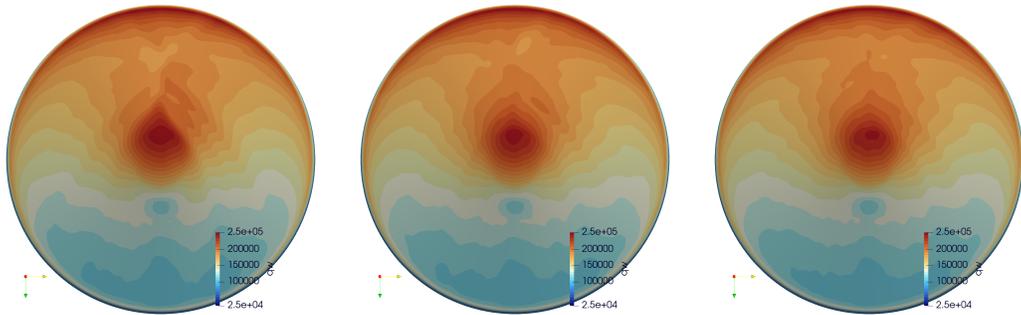

(a) Surface heat flux contours for the initial base solution.
(b) Surface heat flux contours for the first adaptation iteration.
(c) Surface heat flux contours for the second adaptation iteration.
(d) Surface heat flux contours for the third adaptation iteration.
(e) Surface heat flux contours for the fourth adaptation iteration.
(f) Surface heat flux contours for the fifth adaptation iteration.

**Fig. 10  Surface heat flux comparisons between the adaptation iterations.**

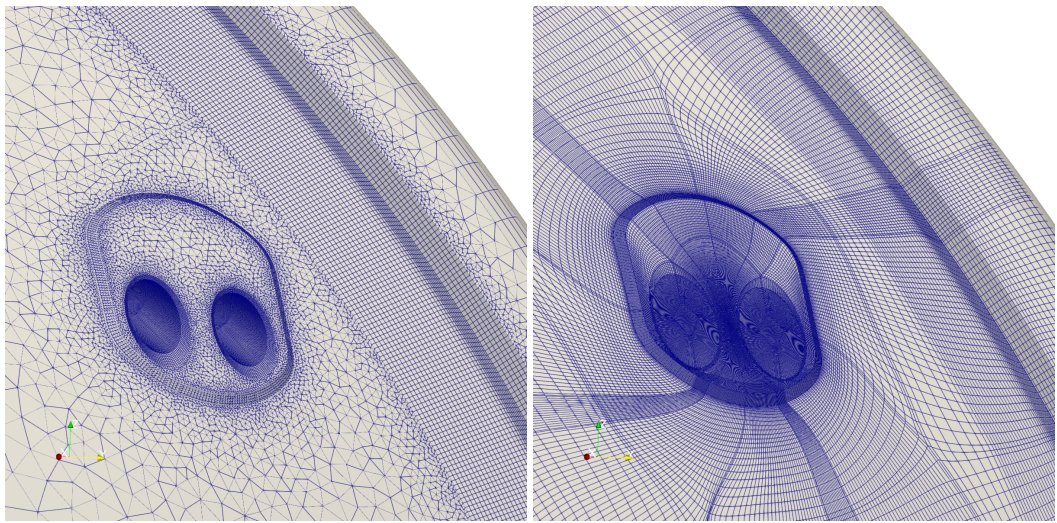

(a) Detail of the surface mesh near the RCS jets on the backshell where the RCS thruster geometry is taken into account.
(b) Detail of the surface mesh for a comparable DPLR simulation where a smooth Outer Mold Line (OML) is considered.

**Fig. 11  Comparing mesh topology near the flush-mounted RCS thruster ports.**

Figure 12b shows a comparison between surface heat flux on the forebody calculated using DPLR and US3D. For the DPLR simulation, half of the computational domain was considered and a symmetry boundary conditions was employed on the $y = 0$ plane. A structured mesh was generated and no turbulence model employed. Next to that, a two-temperature model was used that uses the NASA Lewis database. The surface mesh for the DPLR simulation and for the US3D simulations are shown on the left and right respectively of Figure 12a. The left half of 12a indicates local mesh refinement along the 45° angle on the forebody that is due to the locally required surface mesh refinement around the flush-mounted thruster ports on the backshell as shown in Figure 11b. Figure 12b shows the corresponding surface heat flux for DPLR (left) and US3D (right). The US3D solution is calculated after the fifth adaptation iteration. Qualitatively, the surface heat flux on the forebody appears to agree reasonably well between solvers. The DPLR simulations appears to predict a lower local minima along the symmetry plane and a higher surface heat flux particularly around the upper shoulder region compared to the US3D simulation. This can be ascribed to the mesh refinement study but this can also be due to the fact that the DPLR simulation only considers half of the computational domain while the US3D simulation considers the full domain. Further investigation is required to study these differences more carefully. The local shock stand-off distance for the US3D simulations seems to be a little closer near that upper shoulder region compared to the DPLR simulation (see Figures 12e-12f). Eventhough, the US3D simulation that employs mesh adaptation predicts a locally shorter shock stand-off distance at the top shoulder, the post-shock temperature contours indicate a faster decrease in temperature locally as well. This seems to indicate that different levels of mesh resolution near the shock have a significant consequence on the estimate surface heat flux.

*4. Local flow features in the RCS cavity region*

Finally, the local environment near the RCS jet fairing is studied in more detail. Comparing Figures 7b 8b and 9b indicates that the enviroment predictions in the wake changes drastically when adapting the mesh further. Figure 14 shows a picture of a cross section along the symmetry plane of one the RCS jets on the backshell. Both the mesh and the temperature contours near the upper RCS jets for $\mathcal{T}_0$, $\mathcal{T}_2$ and $\mathcal{T}_5$ are depicted in Figure 14. The initial base mesh is shown Figure 13a and the structured boundary layer mesh is clearly visible both inside the nozzle and on the backshell surface. The rest of the volume consists of isotropic tetrahedra. The temperature contours indicate minor variations where the temperature decreases once moving further into the RCS jet nozzle. Larger variations in temperatures are observed after the second adaptation iteration but the mesh appears to remain fairly isotropic. Higher temperatures are observed after the fifth adaptation iteration and the mesh illustrates anistropic tetrahedra perpendicular to the direction of increased temperature gradient which is as expected. The current hypothesis is that the flow separation from the upper shoulder gets predicted more accurately. The size of the separation bubble that gets formed behind the backward facing step on the backshell increases with each adaptation iteration. The local velocity decreases and therefore higher temperatures are observed. With this example, we have demonstrated that anisotropic solution informed mesh adaptation is essential when simulating hypersonic flows past complex geometries that include various geomtrical features of different scales. A natural extension of this work would be to leverage the present simulations as initial conditions and activate various RCS jet combinations to generate control moments, such as pitch. This mesh adaptation strategy is then able to improve the simulation resolution near the plume and wake and capture the RCS jet plume-wake interaction. This represents a promising direction for follow-on studies.

## V. Conclusions and future work

The goal of this work is to establish a clear workflow that uses metric-based mesh adaptation in the context of realistic aerothermal analysis and design and to illustrate the ease of mesh generation for configurations with detailed geometric features. MIMIC enables the ability to simultaneously adapt the mesh for complex flow and geometrical features. The results discussed in this paper illustrate that using hexahedra versus prisms in the fixed boundary layer mesh result in improved surface heat flux predictions. Furthermore, the results regarding hypersonic flow past the 70° sphere-cone entry capsule illustrate that automated metric-based mesh adaptation is essential to simulate these flows past complex geometries. First, the adaptation was performed based on a flow solution that was calculated using first-order inviscid fluxes in order to avoid severe carbuncle effects. This was repeated for the second adaptation iteration as well. The mesh seemed to be refined sufficiently in the vicinity of the bow shock that no strong carbuncle effects were observed afterwards and a second-order inviscid flux implementation was employed for the following adaptation iterations. Overall, no significant variations in surface heat flux are observed after the third adaptation iteration. The comparison between the US3D simulations that employ MIMIC and the block structured DPLR simulation highlight two things: First, they show that the mesh generation process is significantly simplified. There is no need to generate a single

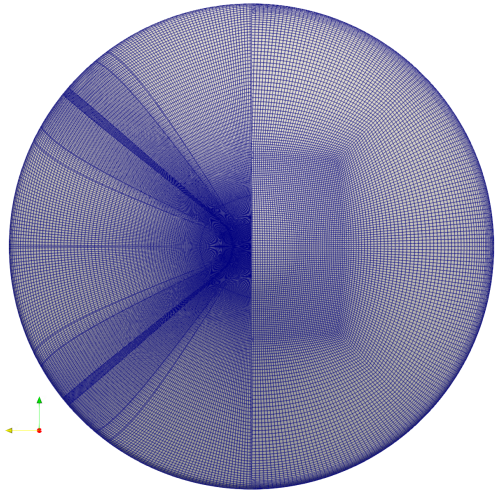
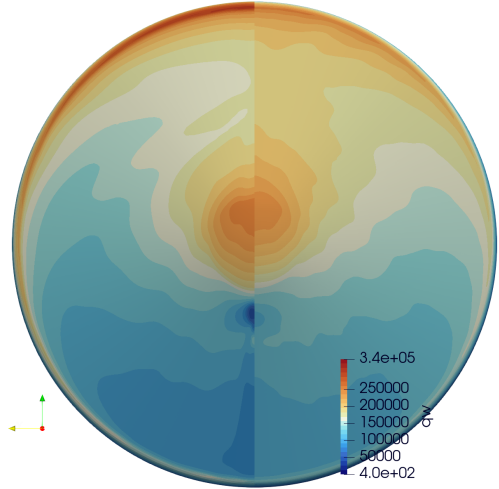

(a) DPLR surface mesh on the left and the US3D surface mesh on the right.

(b) DPLR surface heat flux contours on the left and surface heat flux contours computed using US3D after five adaptation iterations on the right.

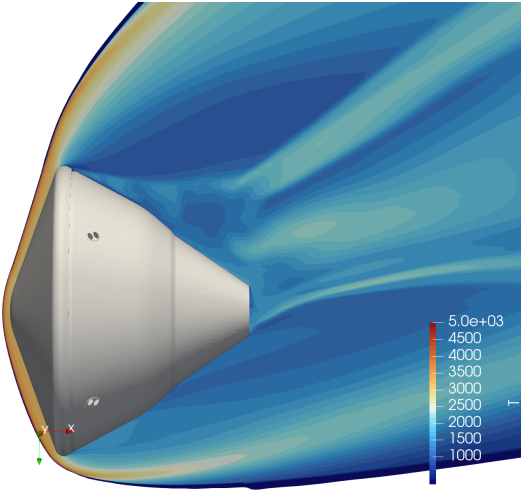
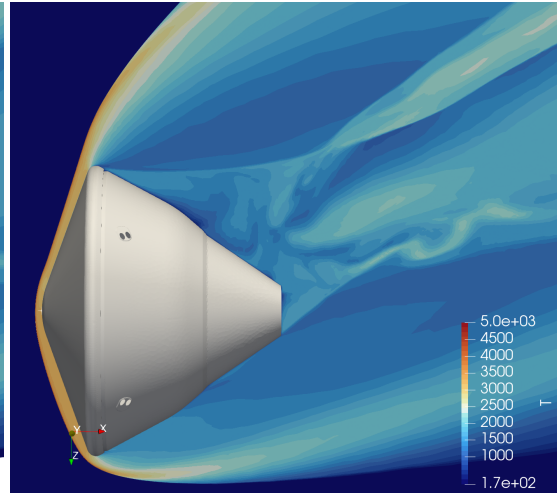

(c) Temperature contours calculated using DPLR on a structured mesh.

(d) Temperature contours calculated using US3D after five adaptation iterations.

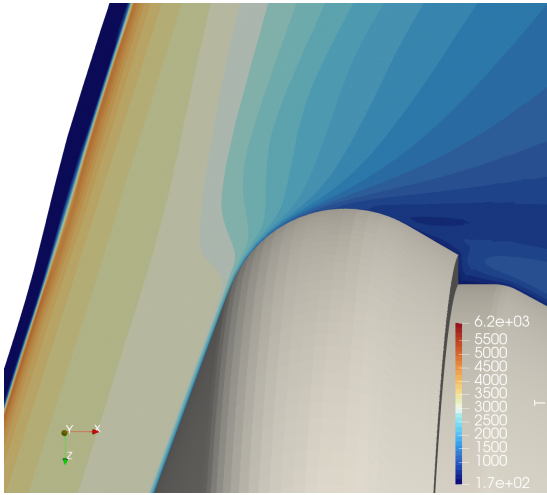
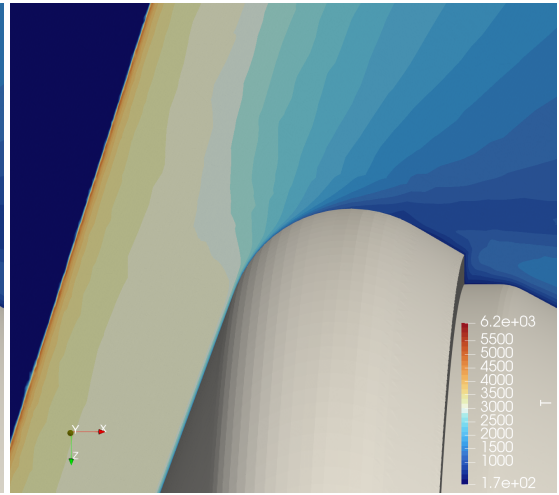

(e) Detail of the temperature contours calculated using DPLR on a structured mesh near the top shoulder.

(f) Detail of the temperature contours calculated using US3D near the top shoulder.

Fig. 12   Comparisons between DPLR and US3D.

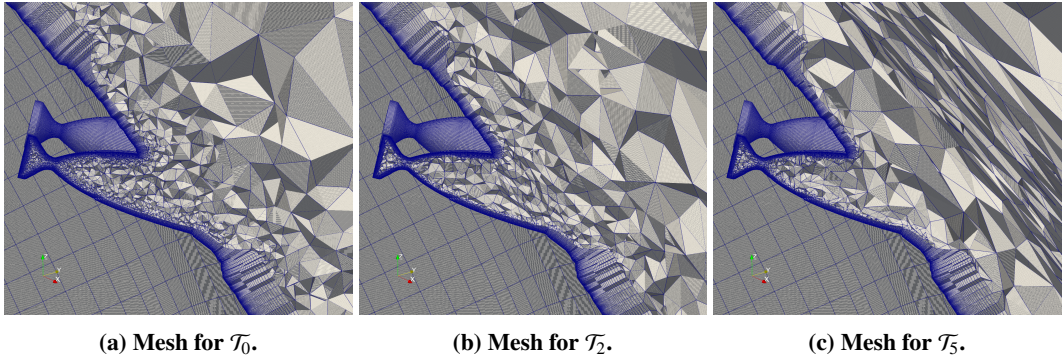

(a) Mesh for $\mathcal{T}_0$.  (b) Mesh for $\mathcal{T}_2$.  (c) Mesh for $\mathcal{T}_5$.

**Fig. 13**

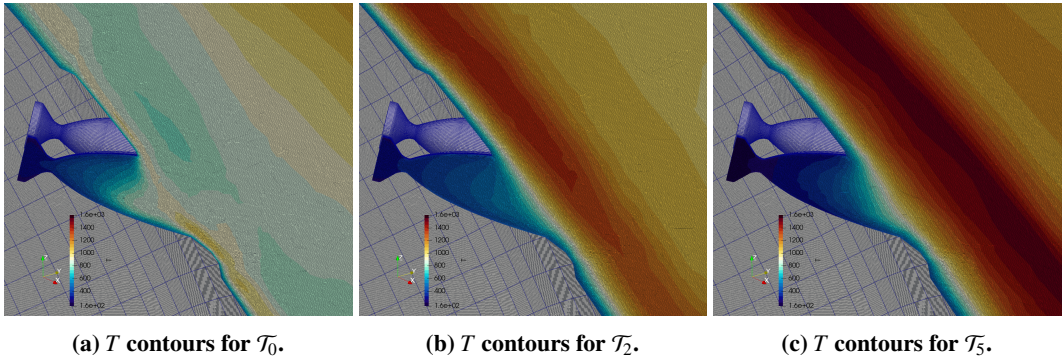

(a) $T$ contours for $\mathcal{T}_0$.  (b) $T$ contours for $\mathcal{T}_2$.  (c) $T$ contours for $\mathcal{T}_5$.

**Fig. 14  Mesh and temperature contours along the symmetry plane of the upper RCS jet on the backshell.**

block structured mesh which is labor intensive and forces the engineer to simplify the geometry. Unstructured adaptive simulations now allows the engineer to run simulations for highly complex geometries that contain geometric features of various scales. Second, this comparison shows that similar levels of accuracy can be obtained when predicting the surface heat flux once significant mesh refinement is applied in the vicinity of the bow shock. For now, no active RCS jet inflow has been considered and the results in this paper solely served as a demonstration that we can now potentially tackle these problems in a monolythic unstructured aerothermal simulation. Futhermore, there is a significant effort to improve numerical schemes that perform better on anisotropic unstructured meshes compared the classical finite volume scheme when it comes to predicting the heat flux. One of these examples is the HyperSolve code [7, 8] which is a hyperbolic Navier-Stokes solver. MIMIC is now able to run in conjuction with HyperSolve and running similar scenarios as the ones presented in this paper with MIMIC and HyperSolve is left for future work.

## VI. Acknowledgments

The author gratefully acknowledges Dr. Dinesh Prabhu for providing the DPLR reference simulation data for both the examples shown in this paper.